\documentclass[twocolumn,showpacs,pra]{revtex4}
\usepackage{amsfonts}
\usepackage{amsmath}
\usepackage{amssymb}
\usepackage{graphicx}

\input{tcilatex}

\begin{document}

\title{Properties of a coupled two species atom-heteronuclear molecule
condensate}
\author{Lu Zhou$^{1,2,3,4}$, Weiping Zhang$^{1,\dag}$, Hong Y. Ling$^{1,5}$,
Lei Jiang$^{6}$ and Han Pu$^{6}$}
\affiliation{$^{1}$Key Laboratory of Optical and Magnetic Resonance Spectroscopy
(Ministry of Education), Department of Physics, East China Normal
University, Shanghai 200062, China}
\affiliation{$^{2}$State Key Laboratory of Magnetic Resonance and Atomic and Molecular
Physics, Wuhan Institute of Physics and Mathematics, Chinese Academy of
Sciences, Wuhan 430071, China}
\affiliation{$^{3}$Center for Cold Atom Physics, Chinese Academy of Sciences, Wuhan
430071, China}
\affiliation{$^{4}$Graduate School, Chinese Academy of Sciences, Beijing 100080, China}
\affiliation{$^{5}$Department of Physics and Astronomy, Rowan University, Glassboro,
New-Jersey, 08028-1700, USA}
\affiliation{$^{6}$Department of Physics and Astronomy, and Rice Quantum Institute, Rice
University, Houston, TX 77251-1892, USA}

\begin{abstract}
We study the coherent association of a two species atomic condensate into a
condensate of heteronuclear diatomic molecules, using both a semiclassical
treatment and a quantum mechanical approach. The differences and connections
between the two approaches are examined. We show that, in this coupled
nonlinear atom-molecule\ system, the population difference between the two
atomic species play significant roles in the ground\textbf{\ }state
stability properties as well as in coherent population oscillation dynamics.
\end{abstract}

\maketitle

\section{Introduction}

After the experimental realization of the trapped atomic Bose-Einstein
condensates (BECs), achieving molecular BEC has been regarded as another
milestone in the field of ultracold atomic physics. As molecules are
inherently much more complex in energy spectrum than their
constitutes-atoms, direct laser cooling methods popular with atoms are
ineffective with molecules. Much recent activities, both in experiments \cite%
{szsrk04,igotbj04,oohesb06,kssbd04,hkbb04,wqs04,wieman} and in theory \cite%
{dkh98,lps04,kvv01,mkj00,ummm05,nmm06,icn04,vya01,wc02,jkn05,stfl06,obc04,jm99,sl03,heizen,tonel}%
, have been focused primarily on converting ultracold atoms into ultracold
molecules by means of magneto- (Feshbach resonance) or photo-association, in
which two atoms are combined into a diatomic molecule mediated by either a
magnetic field or an optical field. Both ultracold degenerate bosonic and
fermionic atoms have been successfully converted into molecules.
Considerable theoretical efforts have been devoted to improving the
conversion efficiency \cite{lps04,kvv01,mkj00,nmm04,nmm06} and understanding
the molecular association \cite%
{dkh98,ummm05,icn04,vya01,wc02,jkn05,stfl06,obc04,jm99,sl03,heizen,tonel} as
well as the dissociation dynamics \cite{jp05, taka, karen} of the
atom-molecule coupling model.

It needs to be emphasized that most of the aforementioned studies, with the
notable exception of Refs.~\cite{nmm06} and \cite{tonel}, concern
homonuclear molecules. The interest of this paper is, however, the
heteronuclear molecules in the coupled atom-molecule systems with two
different atomic species. As a natural progression, quantum degenerate
heteronuclear molecules are expected to be the next challenge to the atomic
physics community, because heteronuclear molecules possess intriguing
properties that will open up many new avenues of research. For example,
unlike their homonuclear counterpart which are always bosonic, heteronuclear
diatomic molecules can be either bosons or fermions, hence quantum
statistics will play important roles in such systems \cite{nmm06}.
Furthermore, large electric dipole moment can be induced in heteronuclear
molecules with the prospect of creating dipolar superfluid \cite{dst03} and
with potential applications in quantum computing \cite{d02}, quantum
simulation \cite{micheli} and test of fundamental symmetry \cite{kl95}. For
these reason, heteronuclear molecules have recently received much
theoretical and experimental attention. Already, Feshbach resonances have
been observed in various quantum degenerate Bose-Fermi atomic mixtures \cite%
{szsrk04,igotbj04,oohesb06}, and heteronuclear molecules from both
Bose-Fermi and Bose-Bose mixtures have been produced through the
photoassociation technique \cite{kssbd04,hkbb04,wqs04}.


In this paper, we consider, within a three-mode model, a system of bosonic
diatomic heternuclear molecules coupled to its constituent atoms, both types
of which are also assumed to be bosonic. Besides the collisional strengths
and the detuning (bare energy difference between the molecular and the
atomic modes), due to the presence of two types of atoms, we have a new
\textquotedblleft control knob\textquotedblright\ --- the population
imbalance between the two species --- which we shall pay special attention
to. We note in passing that recent experiments on two-component degenerate
Fermi gases with population imbalance \cite{ketterle,randy} have generated
great excitement due to its rich phase diagram with various exotic quantum
phases in which the population imbalance plays a critical role. We will
study our system using both a mean-field semiclassical and a full quantum
mechanical method. The differences as well as the connections between the
two approaches will be examined.

The paper is organized as follows. In Sec. \textrm{II} we present our model
in both full quantum and the mean-field version. In Sec. \textrm{III} we
study the ground state properties and their relevance in creating the
molecules from the atoms by adiabatically sweeping the detuning. The
population dynamics is presented in Sec. \textrm{IV} and finally we conclude
in Sec. \textrm{V}. Our work differs from Refs.~\cite{nmm06} and \cite{tonel}
in the following ways: Ref.~\cite{nmm06} focuses on the quantum statistical
properties of the molecules and does not consider the effect of population
imbalance; while Ref.~\cite{tonel} uses a very different quantum approach
(Bethe ansatz) from ours and does not pay much attention to the
atom-molecule conversion process.

\section{Quantum Model and Mean-field Approximation}

We adopt a simple three-mode model in which we describe our atom-molecule
system with two atomic modes (1 and 2) and one molecular mode ($m$). The
basic assumption here is that the spatial wave functions for these modes are
fixed so that we can associate each mode with an annihilation operator $\hat{%
a}_{i}$ of a particle in mode $i$($=1,2$ and $m$). Similar models have been
extensively used in the studies of condensates in double-well potentials 
\cite{double,milburn,tonel2}, coupled atom-molecule condensates \cite%
{jm99,mkj00,nmm06,icn04,vya01,wc02,jkn05,stfl06,tonel,nmm04}, as well as
spinor condensates \cite{lawpu}.

Within the three-mode approximation, the second quantized Hamiltonian reads 
\begin{equation}
\hat{H}=\delta \,\hat{a}_{m}^{\dagger }\hat{a}_{m}+g\left( \hat{a}%
_{m}^{\dagger }\hat{a}_{1}\hat{a}_{2}+h.c.\right) +\sum_{i,j}\chi _{ij}\hat{a%
}_{i}^{\dagger }\hat{a}_{j}^{\dagger }\hat{a}_{j}\hat{a}_{i}\,,  \label{eq7}
\end{equation}%
where the detuning $\delta $ represents the energy difference between the
molecular and atomic levels which can be tuned by external field, $g$ is the
atom-molecule coupling strength and $\chi _{ij}=\chi _{ji}$ is the $s$-wave
collisional strength between modes $i$ and $j$. Without the collisional
terms our model will reduce to the trilinear Hamiltonian describing the
nondegenerate parametric down-conversion in quantum optics \cite{djb93,ma98}.

There are two obvious constants of motion from Hamiltonian (\ref{eq7}):%
\begin{equation}
\hat{N}=\hat{a}_{1}^{\dagger }\hat{a}_{1}+\hat{a}_{2}^{\dagger }\hat{a}_{2}+2%
\hat{a}_{m}^{\dag }\hat{a}_{m},\;\;\;\hat{D}=\hat{a}_{1}^{\dagger }\hat{a}%
_{1}-\hat{a}_{2}^{\dagger }\hat{a}_{2}\,,  \label{eq11}
\end{equation}%
which account for the total particle number and the number difference
between the two atomic species, respectively. Taking advantage of the
constants of motion, the Hamiltonian (\ref{eq7}) can be simplified as%
\begin{eqnarray}
\hat{H} &=&\frac{G}{\sqrt{2N}}\left( \hat{a}_{m}^{\dag }\hat{a}_{1}\hat{a}%
_{2}+h.c.\right) +\frac{\Lambda G}{4N}\left( \hat{a}_{1}^{\dagger }\hat{a}%
_{1}+\hat{a}_{2}^{\dagger }\hat{a}_{2}\right) ^{2}  \notag \\
&&-\frac{\Delta G}{2}\left( \hat{a}_{1}^{\dagger }\hat{a}_{1}+\hat{a}%
_{2}^{\dagger }\hat{a}_{2}\right) \,,  \label{H1}
\end{eqnarray}%
where we have introduced two dimensionless quantities 
\begin{eqnarray*}
\Lambda &=&N\left( \chi _{11}+\chi _{22}+\chi _{mm}+2\chi _{12}-2\chi
_{m1}-2\chi _{m2}\right) /G\,, \\
\Delta &=&\left[ \delta -\left( D-1\right) \chi _{11}+\left( D+1\right) \chi
_{22}+\left( N-1\right) \chi _{mm}\right. \\
&&\left. -\left( N-D\right) \chi _{m1}-\left( N+D\right) \chi _{m2}\right]
/G\,,
\end{eqnarray*}%
with $G=g\sqrt{2N}$ as the rescaled atom-molecule coupling strength. \ In
writing (\ref{H1}), we have neglected the constant terms proportional to $D$
and $N$.

To complement the quantum study, we develop a semiclassical description of
our system by following the usual mean-field approach, which has proven to
be a powerful tool for the study of Bose-Einstein condensates. As a first
step, we apply the Heisenberg equation to arrive at the operator equation
for $\hat{a}_{i}$ and then replace $\hat{a}_{i}$ with the corresponding
c-number $a_{i}$. Next, we change the equation for $a_{i}$ into the ones for 
$N_{i}$ and $\varphi _{i}$ through the transformation $a_{i}=\sqrt{N_{i}}%
e^{i\varphi _{i}}$, where $N_{i}$ and $\varphi _{i}$ represent the number
and phase of the bosonic field for the particles in species $i$,
respectively. Finally, we take advantage of the existence of the two
conserved quantities $N$ and $D$, and simplify our problem into a one
described by two variables: the normalized population in the two atomic
modes 
\begin{equation*}
x=\left( N_{1}+N_{2}\right) /N\,,
\end{equation*}%
and the phase difference 
\begin{equation*}
\varphi =\varphi _{1}+\varphi _{2}-\varphi _{m}\,.
\end{equation*}%
The equations of motion for $x$ and $\varphi $ can be easily obtained as 
\begin{subequations}
\label{mfe}
\begin{eqnarray}
\frac{dx}{d\tau } &=&-\sqrt{\left( 1-x\right) \left( x^{2}-d^{2}\right) }%
\sin \varphi \,,  \label{dx/dt} \\
\frac{d\varphi }{d\tau } &=&\Delta -\Lambda x-\frac{d^{2}+2x-3x^{2}}{2\sqrt{%
\left( 1-x\right) \left( x^{2}-d^{2}\right) }}\cos \varphi \,,
\label{dfai/dt}
\end{eqnarray}%
where $\tau =Gt$ is the dimensionless time and $d=D/N$ the normalized atomic
population imbalance. Without loss of generality, we will assume a
non-negative $d\in \lbrack 0,1]$.

In the language of Hamiltonian mechanics, $x$ and $\varphi $ form a pair of
canonically conjugate variables satisfying the equations 
\end{subequations}
\begin{equation*}
\frac{dx}{d\tau }=\frac{\partial H}{\partial \varphi }\,,\;\;\;\frac{%
d\varphi }{d\tau }=-\frac{\partial H}{\partial x}\,,
\end{equation*}%
with the dimensionless mean-field Hamiltonian $H$ given by 
\begin{equation}
H=\frac{\Lambda }{2}x^{2}-\Delta x+\sqrt{\left( 1-x\right) \left(
x^{2}-d^{2}\right) }\cos \varphi \,.  \label{ch}
\end{equation}%
We note that if $d=0$, i.e., when the two atomic modes have the same
population, Hamiltonian (\ref{ch}) would have the same form as the
corresponding Hamiltonian describing homonuclear molecule association from a
single atomic mode \cite{jkn05,stfl06}. The quantum mechanical Hamiltonian (%
\ref{H1}) and its semiclassical counterpart (\ref{ch}) serve as the starting
point of our study.

\section{Steady States and Rapid Adiabatic Passage}

Semiclassically, the fixed points $\left( x_{0},\varphi _{0}\right) $ are
the steady-state solutions to Eqs.~(\ref{mfe}), and the ground state
corresponds to the ones that give rise to the smallest energy. Obviously $%
x\in \lbrack d,1]$. For convenience, we also introduce a variable 
\begin{equation*}
y=1-x\,,
\end{equation*}%
which lies in the range of $[0,1-d]$\ and has the physical meaning that $y/2$%
\ represents the normalized molecular population. For clarity, we will
separately discuss the two cases: $\Lambda =0$ and $\Lambda \neq 0$.

\subsection{Case 1: $\Lambda =0$}

In order to illustrate the effect of atomic population imbalance, we first
present the results for $d=0$. The ground state in this case is given by 
\begin{equation*}
\left\{ 
\begin{array}{ll}
y_0 = 1,\;\varphi_0\;\mathrm{undefined}\,, & \mathrm{for}\;\Delta \leq -1 \\ 
y_0 = \frac{1}{9}(\sqrt{\Delta ^{2}+3}-\Delta )^2,\;\varphi_0=\pi \,, & 
\mathrm{for}\;\Delta > -1%
\end{array}
\right.
\end{equation*}%
from which one can see that although $y_0$ is continuous throughout the $%
\Delta$-space, the derivative $dy_0/d\Delta$ has a discontinuous jump at $%
\Delta=-1$. Therefore $\Delta=-1$ represents a critical point that separates
the pure molecule phase ($y_0=1$) from the atom-molecule mixture phase in
the semiclassical theory.

\begin{figure}[tbp]
\includegraphics[width=7cm]{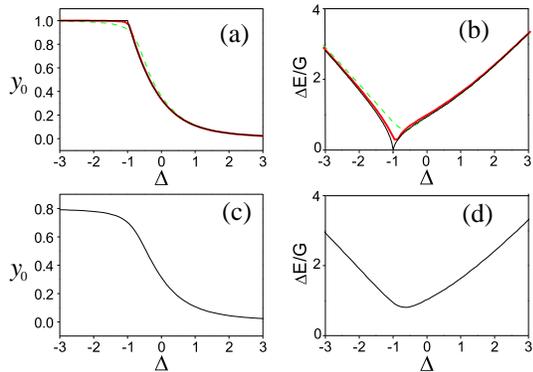}
\caption{{\protect\footnotesize (Color online) Ground state molecular
population }$y_{0}${\protect\footnotesize \ and energy gap }$\Delta E$%
{\protect\footnotesize \ as functions of }$\Delta ${\protect\footnotesize .
For (a) and (b), }$d=0${\protect\footnotesize , the thin black lines
represent the semiclassical results, dashed green lines are quantum results
for }$N=10${\protect\footnotesize , and thick red lines are quantum result
for }$N=100${\protect\footnotesize . For (c) and (d), }$d=0.2$%
{\protect\footnotesize \ and only the semiclassical results are shown.}}
\label{fig1}
\end{figure}

To study the corresponding quantum behavior and its connection with the
semiclassical approach, we expand the Hamiltonian~(\ref{H1}) using Fock
state basis for a given set of $N$ and $D$ and diagonalize the resulting
Hamiltonian matrix. Both the quantum and the semiclassical results of ground
state molecular population are shown in Fig.~\ref{fig1}(a). The quantum
calculation always results in a smooth $y_{0}$ curve although it also shows
a rapid change from 0 to 1 in a small region near $\Delta =-1$. As expected,
the quantum results approach the semiclassical limit as $N$ increases.

Further insights into the properties of the system can be gained by studying
the excitations above the ground state. The quantum many-body excited states
are obtained in the same manner as above through the diagonalization of the
Hamiltonian matrix. We are particularly interested in the \textquotedblleft
energy gap\textquotedblright , $\Delta E$, defined as the energy difference
between the first excited state and the ground state, which is plotted in
Fig.~\ref{fig1}(b) for several different $N$. The energy gap shows a
minimum, which is always finite, at the value of $\Delta $ around which $%
y_{0}$ rapidly approaches 1. The semiclassical energy gap can be obtained
through the following linearization procedure: Substituting $x=x_{0}+\delta
x $ and $\varphi =\varphi _{0}+\delta \varphi $ into Eqs.~(\ref{mfe}) where $%
(x_{0},\varphi _{0})$ are the steady-state solution and $(\delta x,\delta
\varphi )$ represent the small fluctuations away from the steady state,
keeping terms up to first order in fluctuations, we have 
\begin{eqnarray}
\frac{d}{d\tau }\delta x &=&-\sqrt{\left( 1-x_{0}\right) \left(
x_{0}^{2}-d^{2}\right) }\cos \varphi _{0}\delta \varphi ,  \notag \\
\frac{d}{d\tau }\delta \varphi &=&\left\{ -\Lambda -\frac{\left(
1-3x_{0}\right) }{\sqrt{\left( 1-x_{0}\right) \left( x_{0}^{2}-d^{2}\right) }%
}\cos \varphi _{0}\right.  \notag \\
&&\left. +\frac{\left( d^{2}+2x_{0}-3x_{0}^{2}\right) ^{2}}{4\left[ \left(
1-x_{0}\right) \left( x_{0}^{2}-d^{2}\right) \right] ^{3/2}}\cos \varphi
_{0}\right\} \delta x\,,  \label{eq23}
\end{eqnarray}%
where, in anticipation of later studies, we have not made the assumption of $%
\Lambda =0$. The oscillation frequency of $\delta x$ and $\delta \varphi $
can be derived straightforwardly as 
\begin{eqnarray}
\omega ^{2} &=&\left[ \frac{\left( d^{2}+2x_{0}-3x_{0}^{2}\right) ^{2}}{%
4\left( 1-x_{0}\right) \left( x_{0}^{2}-d^{2}\right) }+3x_{0}-1\right] \cos
^{2}\varphi _{0}  \notag \\
&&-\Lambda \sqrt{\left( 1-x_{0}\right) \left( x_{0}^{2}-d^{2}\right) }\cos
\varphi _{0}\,.  \label{eq24}
\end{eqnarray}%
For ground state in the case of $\Lambda =0$, the semiclassical excitation
frequency reduces to 
\begin{equation*}
\omega =\sqrt{\Delta ^{2}+3x_{0}-1}\,,
\end{equation*}%
which is the semiclassical energy gap. In particular, for $d=0$, we have 
\begin{equation*}
\omega =\left\{ 
\begin{array}{ll}
\sqrt{\Delta ^{2}-1}\,, & \mathrm{for}\;\Delta \leq -1 \\ 
\left[ \Delta ^{2}+2-(\sqrt{\Delta ^{2}+3}-\Delta )^{2}/3\right] ^{1/2}\,, & 
\mathrm{for}\;\Delta >-1%
\end{array}%
\right.
\end{equation*}%
which is plotted in Fig.~\ref{fig1}(b). The semiclassical energy gap
vanishes at the critical point $\Delta =-1$ with a discontinuous jump in its
derivative.

Figure \ref{fig1}(a) and (b) clearly show that the quantum result approaches
the semiclassical limit as $N \rightarrow \infty$ and hence the much simpler
semiclassical theory is reliable for large $N$. Furthermore, there is a
critical point at $\Delta=-1$ for $d=0$ in the semiclassical theory which is
absent in the quantum calculations with finite $N$, indicating the fact that
no true quantum phase transition can occur in a finite system.

We now discuss the case with finite atomic population imbalance, i.e., $d
\neq 0$. Although semiclassical solutions to ground state population and
excitation can be obtained analytically in the same fashion as in the
previous case for $d=0$, the expressions are generally too messy to be
instructive. We therefore simply display the results in Fig.~\ref{fig1}(c)
and (d). Again we find that the semiclassical calculation reproduces the
quantum result (not shown in the figure) in the large $N$ limit. One major
difference between $d \neq 0$ and $d=0$ is that in the former there is no
quantum phase transition even in the semiclassical limit: both the
population and the energy gap changes smoothly as $\Delta$ varies, and the
energy gap never becomes zero.

From Fig.~\ref{fig1}, we can also see that starting from a pure two species
atomic condensate, we can coherently create molecular condensate using the
method of rapid adiabatic passage, e.g., by tuning $\Delta $ from a large
positive value to a large negative value. Near perfect atom-to-molecule
conversion \cite{footnote} is achieved when $\Delta $ is swept adiabatically 
\cite{adiabaticity} which is confirmed by our numerical calculations.
However, as we demonstrate next, such a smooth conversion of atoms into
molecules by a slow sweeping of $\Delta $ cannot be taken for granted when $%
\Lambda \neq 0$.

\subsection{Case 2: $\Lambda \neq 0$}

\begin{figure}[tbp]
\includegraphics[width=7.5cm]{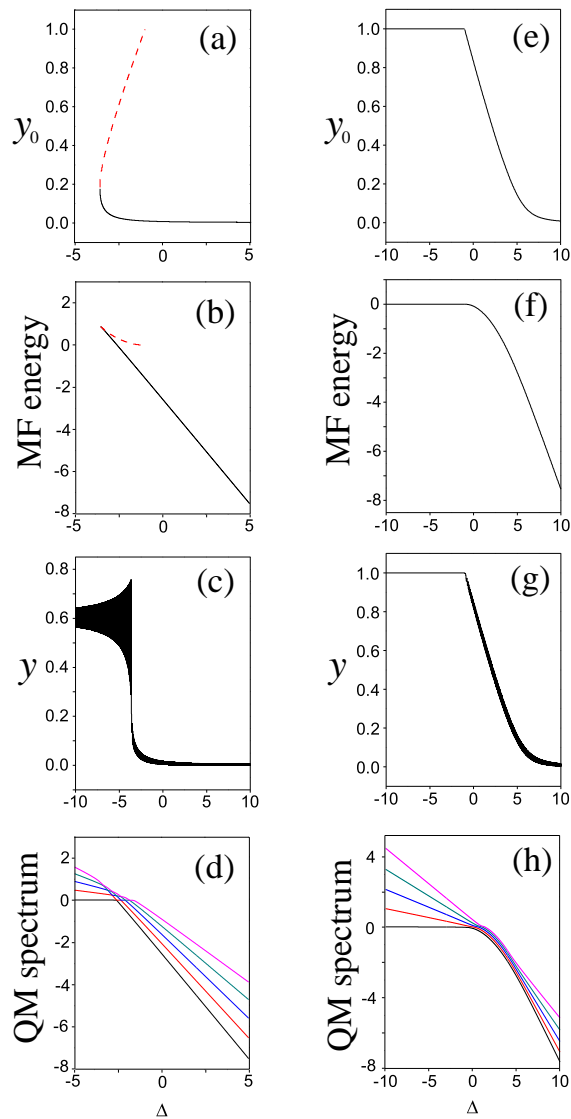}
\caption{{\protect\footnotesize (Color online) Left panel shows an example
of dynamical instability with }$d=0${\protect\footnotesize \ and }$\Lambda
=-5${\protect\footnotesize . (a) Molecular population in the low lying
semiclassical steady states as functions of }$\Delta ${\protect\footnotesize %
; the state represented by the red dashed curve is dynamically unstable. (b)
Corresponding dimensionless semiclassical mean-field energies as calculated
using Eq.~(\protect\ref{ch}) ; (c) Molecular population as }$\Delta $%
{\protect\footnotesize \ is linearly swept; (d) The corresponding quantum
many-body energy spectrum for }$N=20${\protect\footnotesize , only the
lowest five eigenenergies are shown. As the classical Hamiltonian (\protect
\ref{ch}) represents energy per pair of atoms, the quantum eigenenergy (in
units of }$G${\protect\footnotesize ) has been rescaled by a factor of }$%
(N/2)^{-1}${\protect\footnotesize . Right panel is the same as the left
except for }$\Lambda =5${\protect\footnotesize \ where there is no dynamical
instability.}}
\label{fig2}
\end{figure}

With a finite $\Lambda $, the algebra becomes much more complicated. We
resort to numerical calculations in this case. Consider first the
semiclassical situation. The left panel of Fig.~\ref{fig2} illustrates the
properties of the system with $d=0$ and $\Lambda =-5$. Fig.~\ref{fig2} (a)
and (b) show the molecular population and mean-field energy for the
semiclassical steady states. In the region $\Delta \in \lbrack -3.56,-1]$,
there exist three steady states with similar energies as shown in the figure 
\cite{note}. The mean-field energy exhibits a swallowtail loop structure.
Similar structures have been observed in condensates moving in optical
lattice potentials \cite{tail1} and in two-component condensates \cite{tail2}
under certain conditions, and are associated with dynamical instability.

In our system, by calculating the excitation frequency using Eqs.~(\ref{eq23}%
) and (\ref{eq24}), we find that one of the three steady states, represented
by the red dashed lines in Fig.~\ref{fig2} (a) and (b), possesses imaginary
excitation frequency, a signature of dynamical instability. This unstable
state links the two stable ones, representing a classical example of
bistability which has been intensely studied in the context of nonlinear
optics and laser theory \cite{note2}. The existence of such a state is the
key to the development of atom-molecule switch, the matter wave analog \cite%
{jan} of optical bistable switch, for controlling matter waves by matter
waves in a coherent and bistable fashion. Under such a bistable situation,
no matter how slow we tune $\Delta $, the system will not be able to follow
the ground state --- when we enter the dynamical unstable region, a
discontinuous jump will necessarily occur and the atom-molecule conversion
efficiency will suffer. This is confirmed in our numerical simulation as
shown in Fig.~\ref{fig2}(c) where we linearly sweep $\Delta $ from a large
positive to a large negative value starting from a pure two species atomic
condensate. In this example, only about 60\% of the initial atoms will
associate into molecules.

It is instructive to examine the situation from the quantum many-body point
of view. Fig.~\ref{fig2}(d) shows the five lowest eigenenergies of the
quantum Hamiltonian (\ref{H1}) for $N=20$. The quantum mechanical energy
spectrum exhibits a net of anticrossings enveloped by a swallowtail loop
structure that will morph into the semiclassical energy diagram as shown in
Fig.~\ref{fig2}(b). Similar semiclassical-quantum correspondence was
observed in two-component condensates \cite{loop1, loop2} and condensates in
double-well potentials \cite{milburn,tonel2}.

In comparison, the right panel of Fig.~\ref{fig2} shows a situation without
dynamical instability. In this case, rapid adiabatic passage results in a
near perfect atom-molecule conversion, and the system follows the ground
state closely as $\Delta $ is tuned.

\begin{figure}[tbp]
\includegraphics[width=7cm]{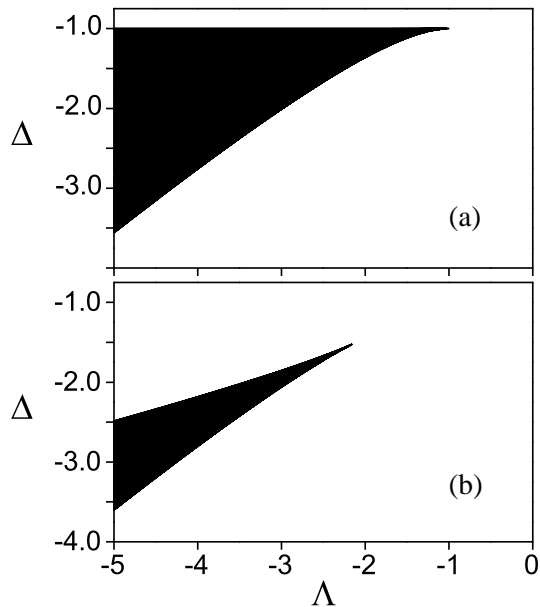}
\caption{{\protect\footnotesize Stability phase diagram in the }$\Lambda $%
{\protect\footnotesize -}$\Delta ${\protect\footnotesize \ parameter space.
The black region are dynamically unstable. (a) }$d=0${\protect\footnotesize %
; (b) }$d=0.2${\protect\footnotesize .}}
\label{fig3}
\end{figure}

Figure \ref{fig2} shows that in order to create molecular condensate with
high efficiency using the rapid adiabatic passage method, it is of crucial
importance to avoid the unstable regimes \cite{lps04}. Fig.~\ref{fig3} shows
the stability phase diagram in the $\Lambda $-$\Delta$ parameter space. We
find that dynamical instability occurs in the region of $\Lambda <-1$ and $%
\Delta <-1$ and is quite sensitive to atomic population imbalance $d$: With
the increase of $d$, the unstable region shrinks. Therefore tuning the
population imbalance provides us with a handle to control the dynamical
stability of the system.

\section{Coherent Atom-Molecule Population Oscillations}

Coherent population oscillation has been predicted \cite%
{jm99,icn04,vya01,jkn05,stfl06,obc04,heizen,sl03} and experimentally
measured \cite{wieman} in systems of homonuclear molecules coupled to atomic
condensate. Besides proving the phase coherence between atoms and molecules,
a measurement of the oscillation frequency can tell us many properties of
the system such as the molecular binding energy, atom-molecule coupling
strength, etc. We therefore want to study in this section the population
oscillation dynamics in our system starting from a pure atomic cloud,
focusing again on the effect of atomic population imbalance.

In a dissipationless system, the total energy is conserved so that the
Hamiltonian represents another constant of motion and the semiclassical
problem becomes integrable. For an initial state with pure atoms, i.e., $x=1$%
, the energy constant according to Eq.~(\ref{ch}) is $E=-\Delta +\Lambda /2$%
. By inserting 
\begin{equation}
\cos \varphi =\frac{\Delta x-\frac{\Lambda }{2}x^{2}+E}{\sqrt{\left(
1-x\right) \left( x^{2}-d^{2}\right) }}\,,  \label{cos}
\end{equation}%
which is obtained from Eq. (\ref{ch}), into Eqs. (\ref{mfe}), we can easily
find that 
\begin{eqnarray}
\left( \frac{dy}{d\tau }\right) ^{2} &=&y\left[ 1-d^{2}-\left( \Delta
^{\prime 2}+2\right) y\right.  \notag \\
&&\left. +\left( 1-\Delta ^{\prime }\Lambda \right) y^{2}-\Lambda
^{2}y^{3}/4 \right] \,,  \label{dy/dt}
\end{eqnarray}%
where $\Delta ^{\prime }=\Delta -\Lambda $ and $y= 1-x$ as before.

The solution to Eq. (\ref{dy/dt}) can be expressed in terms of the
elliptical functions and strongly depends on the roots of the cubic
equations inside the square bracket. A discussion of the solution for the
model with homonuclear molecules $\left( d=0\right) $ can be found in Refs.~%
\cite{icn04,sl03}. Here, in order to gain physical insight into the effect
of the population imbalance on the oscillation dynamics, we will focus on
the simpler case with $\Lambda =0$. Under this condition, Eq.~(\ref{dy/dt})
reduces to%
\begin{equation*}
\left( \frac{dy}{d\tau }\right) ^{2}=y\left[ \left( 1-y\right) ^{2}-d^{2}%
\right] -\frac{1}{4}\Delta ^{2}y^{2}\,,
\end{equation*}%
whose solution, when expressed in terms of Jacobi's elliptic function, has
the form%
\begin{equation}
y=y_{-}\,\text{sn}^{2}\left( \sqrt{y_{+}}\,\tau /2,\sqrt{y_{-}/y_{+}}\right)
\,,  \label{y}
\end{equation}%
where 
\begin{eqnarray}
y_{-} &=&\frac{1}{2}\frac{1-d^{2}}{1+\frac{\Delta ^{2}}{4}+\sqrt{%
d^{2}-1+\left( 1+\frac{\Delta ^{2}}{4}\right) ^{2}}},  \notag \\
y_{+} &=&\frac{1-d^{2}}{4y_{-}}.  \label{ypm}
\end{eqnarray}

Equation (\ref{y}) describes an undamped oscillation in which $y$ changes
from 0 to the peak value $y_{-}$ with a period%
\begin{equation}
T=\frac{4F\left( \frac{\pi }{2},\sqrt{y_{-}/y_{+}}\right) }{\sqrt{y_{+}}}\,,
\label{t}
\end{equation}%
where $F\left( \pi /2,k\right) $ is the complete elliptic integral of the
first kind.

\begin{figure}[tbh]
\includegraphics[width=7cm]{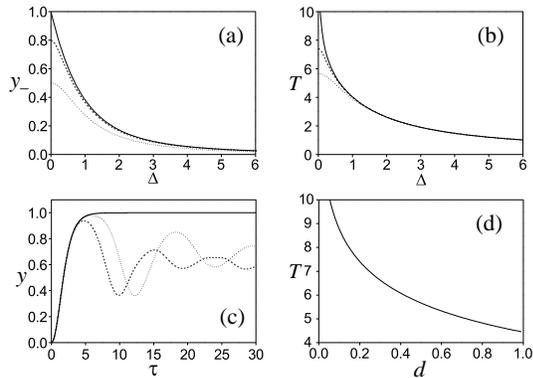}
\caption{{\protect\footnotesize \ (a) and (b), molecular population
oscillation amplitude and period, respectively. The three curves correspond
to }$d=0${\protect\footnotesize , 0.2 and 0.5 in descending order. (c)
Molecular population dynamics for }$\Lambda =\Delta =d=0$%
{\protect\footnotesize . Solid line: semiclassical result; dashed and dotted
lines: quantum result for }$N=100${\protect\footnotesize \ and }$N=1000$%
{\protect\footnotesize , respectively. (d) On resonance semiclassical
oscillation period as a function of }$d${\protect\footnotesize .}}
\label{fig4}
\end{figure}

We plot the amplitude $y_{-}$ and period $T$ of the molecular population
oscillation with respect to $\Delta $ for different $d$ in Fig.~\ref{fig4}%
(a) and (b), respectively. The figure is symmetric with respect to $\Delta
=0 $ so we only present the case with $\Delta \geq 0$. From Eq. (\ref{ypm}),
we find that for any given $d$, the oscillation reaches a maximum value of 
\begin{equation*}
y_{-}=1-d\,,
\end{equation*}%
at resonance, i.e., $\Delta =0$.

One peculiarity from the semiclassical calculation is that when $d=0$, the
oscillation period diverges at $\Delta =0$. In this case we have $%
y_{+}=y_{-}=1$ and Eq.~(\ref{y}) becomes 
\begin{equation*}
y=\text{tanh}^{2}\left( \tau /2\right) \,,
\end{equation*}%
which shows that atomic (molecular) population decreases (increases)
monotonically until all the atoms are converted to molecules. The quantum
mechanical calculation, however, does show damped population oscillations
under the same condition, as illustrated in Fig.~\ref{fig4}(c). The
difference between the semiclassical and the quantum results arises because
the former does not take atom-molecule entanglement into account. The same
behavior will also occur in homonuclear molecule association and has been
studied in Ref.~\cite{vya01}. In heteronuclear molecule association with
finite $d$, the period $T$ as given by Eq.~(\ref{t}) never diverges. Using
the asymptotic formula for $F(\pi /2,k)$, one can show that, on resonance, 
\begin{equation*}
T\approx \frac{2}{\sqrt{1+d}}\,\ln \frac{16}{d}\,,
\end{equation*}%
for small $d$. The resonant oscillation period as a function of $d$ is shown
in Fig.~\ref{fig4}(d).

The situation becomes much more complicated in the case of $\Lambda \neq 0$
and in general no simple analytic formula for population oscillation can be
found. The general features are nevertheless still preserved: semiclassical
result shows undamped oscillation while quantum calculation yields damped
oscillation, and the quantum result approaches the semiclassical limit as $N$
increases.

\bigskip

\section{Conclusion}

In conclusion, we have studied coherent association of a two species atomic
condensate into heteronuclear molecular condensate using a three-mode model,
emphasizing the effect of atomic population imbalance. In particular, the
population imbalance, together with detuning and collisional interaction
strength, will significantly affect the excitation and stability properties
as well as coherent population oscillations of the system. We have also
carefully analyzed the differences and connections between the semiclassical
and the quantum many-body treatments.

\acknowledgments This work is supported by the National Natural Science
Foundation of China under Grant No. 10474055 and No. 10588402, the National
Basic Research Program of China (973 Program) under Grant No. 2006CB921104,
the Science and Technology Commission of Shanghai Municipality under Grant
No. 05PJ14038, No. 06JC14026 and No. 04DZ14009 (WZ), and by the US National
Science Foundation (HP and HYL), and the US Army Research Office (HYL).

$\dag $ To whom correspondence should be addressed E-mail:
wpzhang@phy.ecnu.edu.cn


\begin{thebibliography}{99}
\bibitem{footnote} One can measure the atom-molecule conversion efficiency
by the final molecular population, $y$. In the presence of atomic population
imbalance, the maximum possible value for $y$ is $(1-d)$. Therefore one can
define $y/(1-d)$ as the conversion efficiency.
\bibitem{note} Except for the steady state with $y_{0}=1$ whose relative
phase $\varphi $ is undefined, $\varphi =\pi $ for the other two steady
states. There also exsit one or more steady states with $\varphi =0$. These
states have higher energies and are not shown in the figure.
\end{thebibliography}
\end{document}